\documentclass{aastex}
\usepackage{emulateapj5}

\newcommand{\src}{4U~1728$-$34}

\newcommand{\xte}{{\it RXTE}}
\newcommand{\fpk}{$F_{\rm peak}$}
\newcommand{\fpkre}{$F_{\rm peak,RE}$}

\newcommand{\eps}{{\rm erg\,s^{-1}}}
\newcommand{\epcs}{{\rm erg\,cm^{-2}\,s^{-1}}}
\newcommand{\cts}{{\rm count\,s^{-1}}}

\newcommand{\pasm}{P_{\rm ASM}}

\newcommand{\burstnum}{81}
\newcommand{\preburst}{66}
\newcommand{\preburstmaybe}{8}
\newcommand{\nonpreburst}{15}
\newcommand{\fmean}{$9.2\times10^{-8}\ \epcs$}
\newcommand{\fnopremean}{$4.5\times10^{-8}\ \epcs$}
\newcommand{\fstddev}{$8.7\times10^{-9}\ \epcs$}
\newcommand{\fstddevp}{9.4}
\newcommand{\netvar}{$(4.2)\times10^{-8}\ \epcs$}
\newcommand{\netvarp}{46}

\newcommand{\lastpublic}{2001 November 15}

\slugcomment{Submitted to \apj; revised version v.1.15, \today}

\shortauthors{Galloway et al.}
\shorttitle{Eddington-limited X-ray Bursts from \src}

\begin{document}

\title{Eddington-limited X-ray Bursts as Distance Indicators.\\ I. 
Systematic Trends and Spherical Symmetry in Bursts from \src\\}

\author{Duncan K. Galloway\altaffilmark{1}, Dimitrios Psaltis\altaffilmark{1,2},
  Deepto Chakrabarty\altaffilmark{1,3,4}, and Michael P. Muno\altaffilmark{1,3}}
\altaffiltext{1}{\footnotesize 
  Center for Space Research,
  Massachusetts Institute of Technology, Cambridge, MA~02139}
\altaffiltext{2}{Current address: School of Natural Sciences, Institute
  for Advanced Study, Einstein Dr., Princeton, NJ 08540}
\altaffiltext{3}{Department of Physics, Massachusetts Institute of
Technology}
\altaffiltext{4}{Alfred P. Sloan Research Fellow}
\email{duncan, deepto, muno@space.mit.edu; dpsaltis@ias.edu}

\begin{abstract}
We investigate the limitations of thermonuclear X-ray bursts as a distance
indicator for the weakly-magnetized accreting neutron star 4U~1728$-$34.
We measured the unabsorbed peak flux of \burstnum\ bursts in public data
from the {\em Rossi X-Ray Timing Explorer}.  The distribution of peak
fluxes was bimodal: \preburst\ bursts exhibited photospheric radius
expansion (presumably reaching the local Eddington limit) and were
distributed about a mean bolometric flux of \fmean, while the remaining
(non-radius expansion) bursts reached \fnopremean, on average.
The peak fluxes of the radius-expansion bursts were not constant,
exhibiting a standard deviation of \fstddevp\% and a total variation of
\netvarp\%.  These bursts showed significant correlations between their
peak flux and the X-ray colors of the persistent emission immediately
prior to the burst.  We also found evidence for quasi-periodic variation
of the peak fluxes of radius-expansion bursts, with a time scale of
$\simeq 40$~d.  The persistent flux observed with \xte/ASM over 5.8~yr
exhibited quasi-periodic variability on a similar time scale.  We suggest
that these variations may have a common origin in reflection from a warped
accretion disk.  Once the systematic variation of the peak burst fluxes is
subtracted, the residual scatter is only $\simeq 3$\%, roughly consistent
with the measurement uncertainties.  The narrowness of this distribution
strongly suggests that i) the radiation from the neutron star atmosphere
during radius-expansion episodes is nearly spherically symmetric, and ii)
the radius-expansion bursts reach a common peak flux which may be
interpreted as a standard candle intensity.
Adopting the minimum peak flux for the radius-expansion bursts as the
Eddington flux limit, we derive
a distance for the source of 4.4--4.8~kpc (assuming
$R_{\rm NS}=10$~km), with the uncertainty arising from the probable range
of the neutron star mass
 $M_{\rm NS}=1.4$--2~$M_\sun$.
\end{abstract}


\section{INTRODUCTION}

Thermonuclear (type I) X-ray bursts manifest as rapid changes in the X-ray
intensity of accreting neutron stars in low-mass X-ray binary (LMXB)
systems, with rise times between $\la 1-10$~s and decay times between
$\sim 10-100$~s \cite[see][for reviews]{lew93,bil98a}. Such bursts have
been observed from more than 70 sources \cite[]{lmxb01} and are caused by
unstable nuclear burning of accreted matter on the neutron-star surface.

If the thermonuclear energy during a burst is released sufficiently
rapidly, the flux through the neutron star atmosphere may reach the local
Eddington limit, at which point the outward radiation force balances
gravity.  The excess energy is converted into potential and kinetic energy
of the X-ray photosphere, which is lifted above the neutron star surface
while the emerging luminosity (measured locally) remains approximately
constant and equal to the Eddington limit.  These are the so-called
radius-expansion or Eddington-limited bursts.

For spherically symmetric emission, the Eddington luminosity measured
by an observer at infinity is given by \cite[]{lew93}
\begin{eqnarray}
  L_{\rm Edd,\infty} & = & \frac{8\pi G m_{\rm p} M_{\rm NS} c
  [1+(\alpha_{\rm T}T_{\rm e})^{0.86}]} {\xi\sigma_{\rm T_{\rm e}}(1+X)}
  \left(1-\frac{2GM_{\rm NS}}{Rc^2}\right)^{1/2} \nonumber \\ & = &
  2.5\times10^{38} \left(\frac{M_{\rm NS}}{M_\odot}\right)
  \frac{1+(\alpha_{\rm T}T_{\rm e})^{0.86}}{\xi(1+X)}\left(1-\frac{2GM_{\rm NS}}{Rc^2}\right)^{1/2}\ \eps
  \label{ledd}
\end{eqnarray}
where $M_{\rm NS}$ is the mass of the neutron star, $T_{\rm e}$ is the
effective temperature of the atmosphere, $\alpha_{\rm T}$ is a coefficient
parametrizing the temperature dependence of the electron scattering
opacity \cite[$\simeq 2.2\times10^{-9}$~K$^{-1}$;][]{lew93}, $X$ is the
mass fraction of hydrogen in the atmosphere ($\approx0.7$ for cosmic
abundances), and the parameter $\xi$ accounts for possible anisotropy of
the burst emission.  The final factor in parentheses represents the
gravitational redshift due to the compact nature of the neutron star, and
also depends upon the height of the emission above the neutron star
surface $R\ge R_{\rm NS}$.  Because the Eddington luminosity depends on
the ratio $M_{\rm NS}/R_{\rm NS}$, measurements of the peak flux of
radius-expansion bursts allows in principle the measurement
of these fundamental properties \cite[e.g.][]{damen90,smale01,kuul01}. On
the other hand, because the masses and radii of stable neutron stars
predicted by any given equation of state span a narrow range of values
\cite[see, e.g.,][]{lp01}, Eddington-limited X-ray bursts can be used as
distance indicators.

The validity of the physical picture of Eddington-limited bursts discussed
above can be verified observationally in two ways. First, the peak fluxes
of such bursts from each individual source should be the same. Second, the
peak luminosities inferred for sources with independent distance
measurements, such as those in globular clusters, should correspond to the
Eddington limit for a neutron star.  Since the original discovery of
Eddington-limited bursts, a number of authors have addressed these
questions. In early observations, a significant number of radius expansion
bursts had been detected from three sources and their peak fluxes were
found to be similar to within $\simeq 20$\% (4U~1820$-$30:
\citealt{vlvp86,damen90};
4U~1636$-$536: \citealt{damen90}; 4U~1728$-$34: \citealt{bas84}). In
particular, the peak luminosities of radius-expansion bursts observed from
4U~1820$-$30, which resides in the gobular cluster NGC~6624 were found to
be comparable to the Eddington limit for a neutron star. Although successful
in providing support to the model of Eddington-limited bursts, these early
studies were limited to a small number of sources and suffered from the
statistical uncertainties inherent to fitting spectral models to low
signal-to-noise data.
In recent years, observations with {\it BeppoSAX}\/ and the {\em Rossi
X-ray Timing Explorer\/} (\xte) have revealed a large number of
Eddington-limited bursts from several sources, which have been studied in
great detail.
Using {\it
BeppoSAX}\/ and \xte\/ data, \cite{kuul02} recently studied the
Eddington-limited bursts of 12 globular cluster sources with well known
distances and showed that, with one exception, their peak fluxes were
constant to within $\simeq 15$\% and were comparable to the Eddington limit
for neutron stars with H-poor atmospheres.

In this series of articles, we use all the publicly
available data obtained with \xte\/ to date
in order to observationally test the hypothesis that Eddington-limited
bursts can be used as distance indicators for neutron-star LMXBs.  In the
present study we quantify, and examine the causes of, systematic variations
in the peak burst flux from the source with the greatest number of bursts
detected by \xte, \src.

\section{RXTE OBSERVATIONS OF \src}

The X-ray source \src\ (GX~354+0; $l=354\fdg3$, $b=-0\fdg15$) was first
resolved by {\it Uhuru}\/ scans of the Galactic center region
\cite[]{ftj76}. Thermonuclear X-ray bursts from \src\ were discovered
during {\em SAS-3}\/ observations
\cite[]{lcd76,hoff76}.  The bursting behavior was subsequently studied in
detail using extensive  observations by {\em SAS-3\/}, which accumulated
96 bursts in total. From these data \cite{bas84} showed evidence for a
narrow distribution of peak burst fluxes, as well as a correlation between
the peak flux and the burst fluence. The distance to the source has previously
been estimated from measurements of the peak burst fluxes as between
4.2--6.4~kpc \cite[]{vp78,bas84,kam89}.  The estimated extinction to the
source is $A_V\approx14$; only a precise position following from the
detection of a radio counterpart allowed identification with a $K=15$
infrared source 
%%%%%%%%%%%%%%%%%%%
\centerline{\epsfxsize=8.5cm\epsfbox{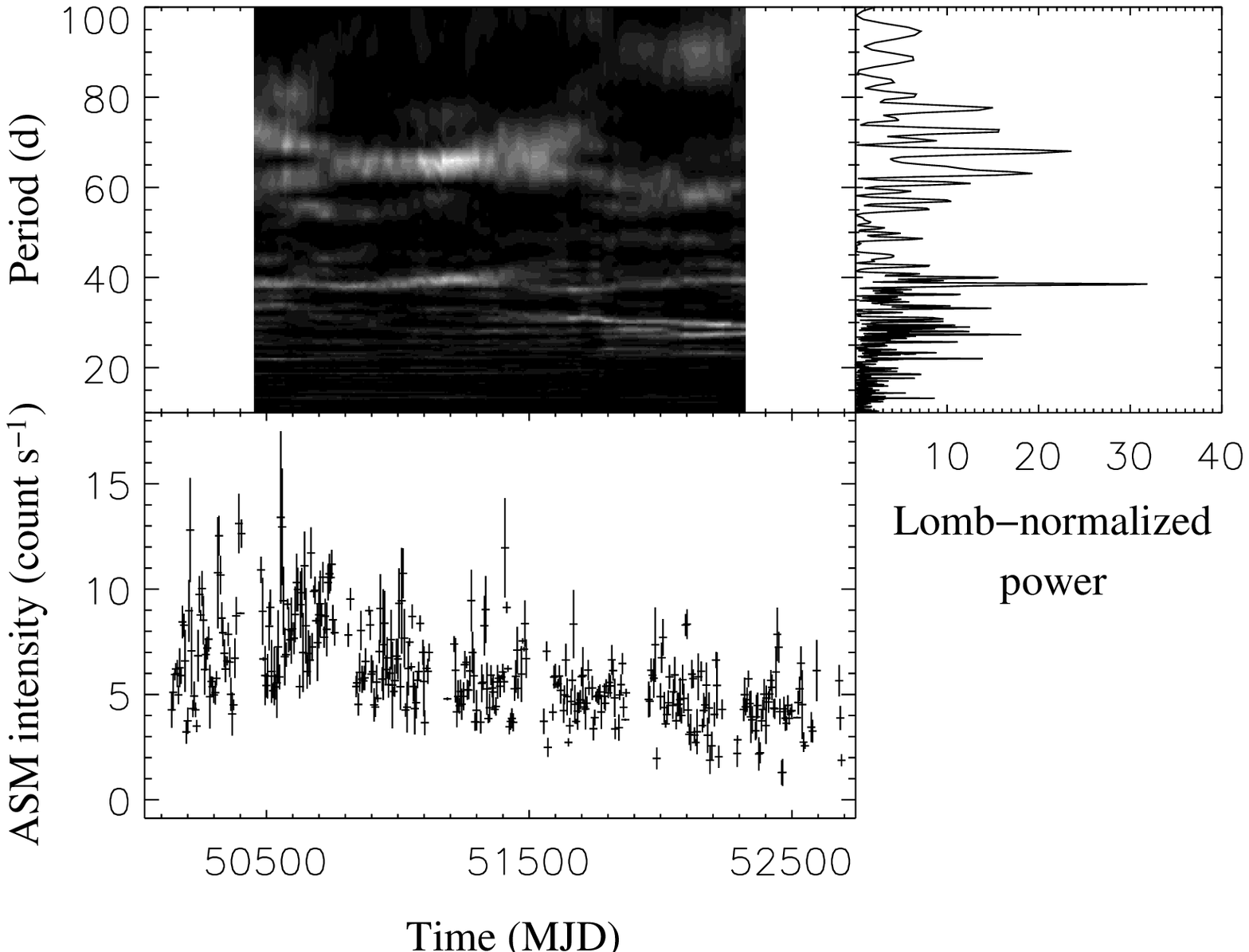}}
 \figcaption{\xte/ASM measurements of \src.
The top right panel shows the Lomb-normalized periodogram computed over the
entire ASM history. The top left panel shows the corresponding dynamical Lomb
periodogram, calculated from 730-d subsets of the ASM data at 9-d
intervals.  The lower panel shows the 5-d averaged count rates, computed
from the initial 1-d averages excluding those points where the measured
error $>0.5\ \cts$.  Error bars, where shown, indicate the $1\sigma$
uncertainties.
 \label{asmdat} }
\bigskip
\noindent
%%%%%%%%%%%%%%%%%%%
\cite[]{marti98}.  No independent distance
measurement is available. Long-term {\it Ariel-5}\/
measurements, as well as the first two years of monitoring by the All-Sky
Monitor (ASM) on board \xte, suggest the presence of a long-term
quasi-periodicity of 63 or 72~d, respectively \cite[]{kong98}. 

\xte/PCA observations of the source in 1996 led to the discovery of nearly
coherent millisecond oscillations during the X-ray bursts
\cite[]{stroh96}.  Similar oscillations were subsequently observed in 9
other sources \cite[]{vdk00,wij01,1916burst,kaaret02}.  A substantial
archive (1140~ks) of public \xte/PCA data from \src\ has accumulated
throughout these observations, dating from shortly after the launch of the
satellite on 1995 December 30.
Subsets of the bursts observed during the PCA observations have been
studied by \cite{vs01} and \cite{franco01}, with particular attention to
the relationship between the appearance of burst oscillations and the mass
accretion rate. A significant fraction of the bursts observed by \xte\/
show evidence for photospheric radius expansion, exhibiting the
characteristic temporary increase in the apparent blackbody radius,
coincident with a decrease in the color temperature in the early stages of
the burst.  \cite{muno01} used this dataset to find a correlation between
the frequency of the burst oscillations and the preferential appearance in
radius expansion or non-radius expansion bursts in nine sources.

We have obtained all the available public \xte\ data to date from the
High-Energy Astrophysics Science Archive Research Center (HEASARC;
{\url http://heasarc.gsfc.nasa.gov}). This study is part of a larger
effort involving more than 70 known bursters; the burst detection and
analysis procedures, as well as the resulting burst database are
described in more detail in Galloway et al. (2003b, in preparation).
Several of the bursts from \src\ we analyzed have also been studied in
detail by
\cite{stroh96,stroh97b,stroh98}.

 \subsection{ASM}

The All-Sky Monitor (ASM) onboard \xte\/ consists of three Scanning Shadow
Cameras (SSCs) sensitive to photons in the energy range 2--10~keV, mounted
on a rotating platform \cite[]{asm96}.  ASM observations are performed as
sequences of dwells lasting 90~s, after which the platform holding the
SSCs is rotated by 6 degrees.  Most of the sky is viewed once every few
hours.  The data from each SSC from each dwell are averaged to obtain the
daily intensities of known sources in the field of view.
The long-term 2--10~keV \xte/ASM flux history of the source is shown in
Fig.~\ref{asmdat}.  The observed photon flux exhibited variations of
$\sim5\ \cts$ on time scales of $\sim10$~d, superimposed on a long-term
trend of decreasing mean intensity.

We computed the Lomb-normalised periodogram \cite[]{pr89} of the full
dataset, shown also in Fig.~\ref{asmdat}.  We found significant evidence
for a periodicity with
$\pasm=38.6$~d, with a Lomb-normalized power
of 31.9 as well as secondary peaks at 63.7 and 67.5~d. The latter values
are close to those found from earlier {\it Ariel-5}\/ measurements, as
well as a 2-yr subset of the \xte/ASM data by \cite{kong98}. The rms
amplitude of the signal at $\pasm$ was $\simeq 8-9$\%. The dynamical
Lomb-normalized periodogram shows that this periodicity
varied in strength over the entire ASM history (Fig.~\ref{asmdat}). The
$\approx39$~d periodicity appeared strongly only in the first half of the
measurement interval, when the source reached its peak long-term
intensity.  After MJD~51400, the dominant periodicity appeared to shift to
around 30~d. A broad peak around 65~d was also present
at most times.

 \subsection{PCA}
\label{pca}

The Proportional Counter Array \cite[PCA;][]{xte96} consists of five
identical gas-filled proportional counter units (PCUs) with a total
effective area of $\approx6000\ {\rm cm}^2$, and sensitivity to X-ray
photons in the 2--60~keV range.  We initially scanned 1~s binned
lightcurves over the full PCA energy range (created from ``Standard-1''
mode data) in order to locate X-ray bursts from \src.  In all the PCA
observations, data were also collected in user-defined modes offering higher
time resolution, compared to the standard data modes.  The PCA records the
arrival time (1~$\mu$s resolution) and energy (256 channel resolution) of
every unrejected photon; data were generally binned on somewhat lower time
and spectral resolution in order to meet telemetry limits.  Once an X-ray
burst was detected, we extracted 2--60~keV PCA spectra within intervals of
0.25--1~s covering the burst.  A response matrix was generated separately
for each burst using {\sc pcarsp} version
8.0 (part of {\sc lheasoft} release 5.2, 2002 June 25) in order to take
into account the known gain variations over the life of the instrument.
The gain was manually re-set by the instrument team on 3 occasions (1996
March 21, 1996 April 15 and 1999 March 22); an additional gain epoch (5)
began on 2000 May 13 with the loss of the propane layer in PCU \#0.  In
addition to these abrupt changes, a more gradual variation in the
instrumental response is known to occur, due to a number of factors.

We then analyzed these data using an approach that is often used in X-ray
burst analysis
(e.g. \citealt{kuul01}, although see also \citealt{vpl86}). 
For the background to the burst spectra we used a spectrum 
%%%%%%%%%%%%%%%%%%%%%%%%%%
\centerline{\epsfxsize=8.5cm\epsfbox{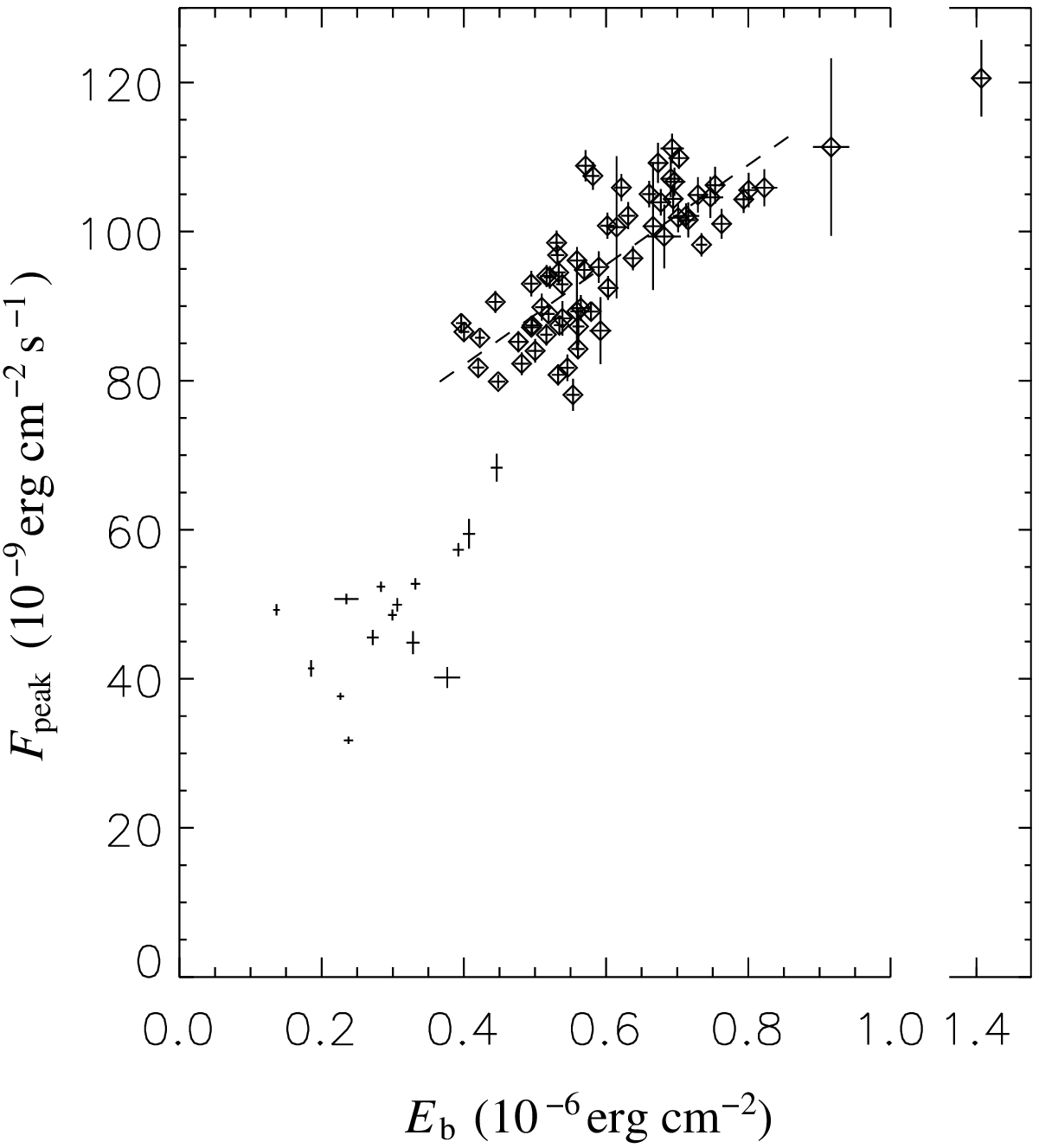}}
\figcaption{
The peak flux \fpk\ (excluding the pre-burst persistent emission) plotted
as a function of the integrated burst fluence $E_{\rm b}$
of \nonpreburst\ non-radius expansion (cross) and \preburst\
radius expansion (diamond) bursts from \src.  The result of a linear fit
to \fpk\ as a function of $E_{\rm b}$ for the radius expansion bursts is
plotted as a dashed line. Error bars indicate the
$1\sigma$ uncertiainties. Note the broken $x$-axis.
  \label{fluxhist} }
\bigskip
\noindent
%%%%%%%%%%%%%%%%%%%%%%%%%%
extracted from
a (typically) 16~s interval prior to the burst.
Each time-resolved background-subtracted spectrum during the bursts
was fit with a blackbody model multiplied by a low-energy cutoff
representing interstellar absorption with fixed abundances. The
initial spectral fitting was performed with the absorption column
density $n_{\rm H}$ free to vary; the resulting fit values typically
exhibited very large scatter, particularly towards the end of the
burst when the flux was low.  Thus, for the final analysis we re-fit
each spectra with $n_{\rm H}$ fixed at the weighted mean value
measured over the entire burst.
The unabsorbed bolometric flux $F_{{\rm bol},i}$ at each timestep $t_i$
was then estimated according to
\begin{eqnarray}
  F_{{\rm bol},i} & = & \sigma T_i^4 \left( \frac{R_{\rm NS}}{d} \right)_i^2 \nonumber \\
      & = & 1.0763\times10^{-11}\ T_{{\rm bb},i}^4 K_{{\rm bb},i}^2\ \epcs
\label{flux}
\end{eqnarray}
where $R_{\rm NS}$ is the neutron star radius, $d$ is the distance to the
source, $T_{\rm bb}$ is the blackbody (color) temperature in units of
keV, and $K_{\rm bb}$ is the normalisation of the blackbody component
as returned by the fitting program ({\sc xspec} version 11). We also
estimated the burst fluence $E_{\rm b}$ by integrating the measured
$F_{{\rm bol},i}$ over the burst duration. We discuss the possible
consequences of the bolometric correction implicit in
equation~(\ref{flux}) in section \S\ref{system}.
\cite{kuul01} and other authors have noted a $\approx20$\% systematic flux
offset in \xte\/ measurements compared to other instruments. It may be
argued that the absolute flux calibration of these instruments is no
better than that of \xte; in any case, the \xte\/ data still offer
substantially better signal-to-noise and hence greater precision of
%%%%%%%%%%%%%%%%%%%%%%%%%%%%%%%
\centerline{\epsfxsize=8.5cm\epsfbox{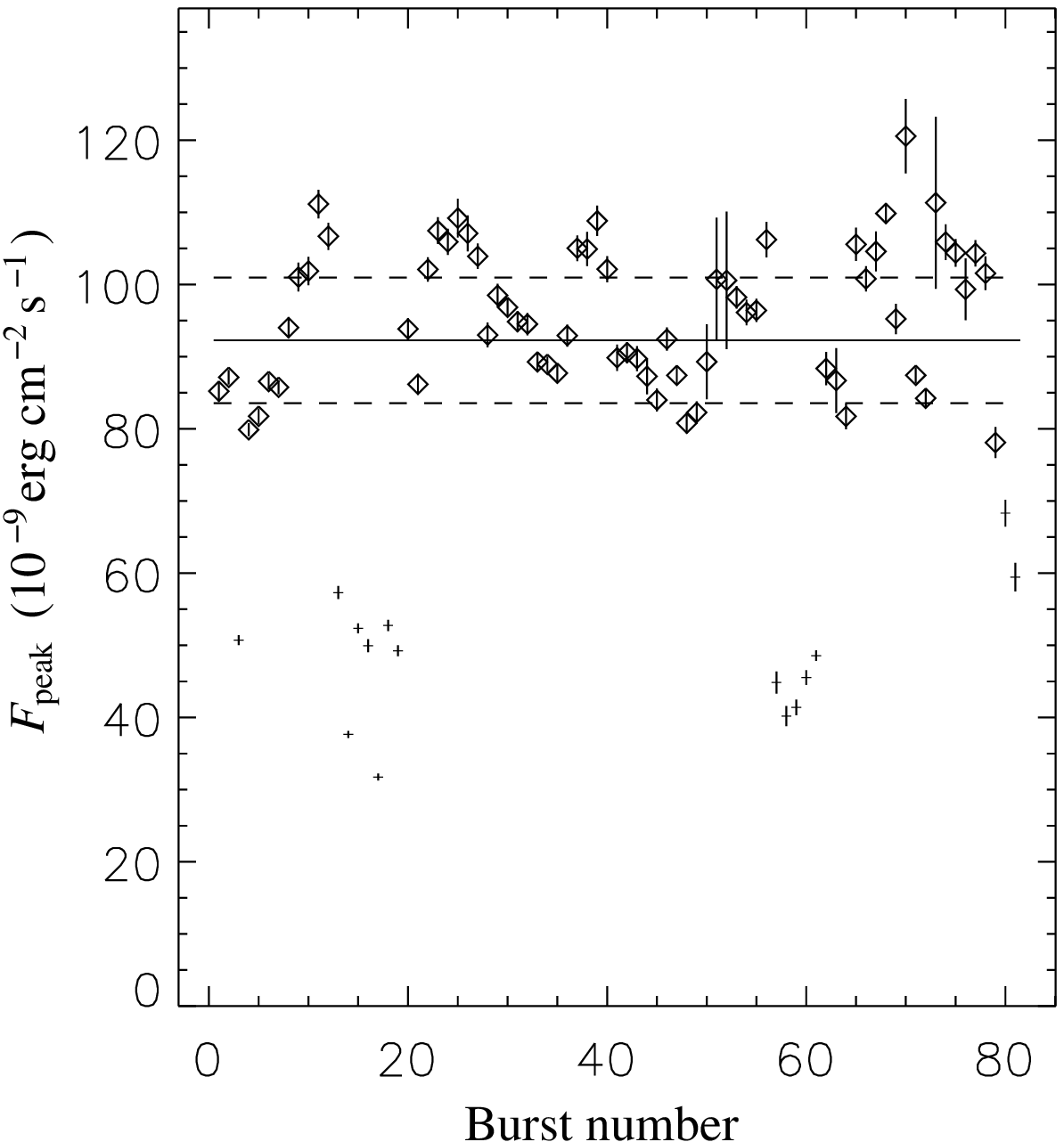}}
\figcaption{
  The peak fluxes \fpk\ (excluding the pre-burst persistent emission) of
\burstnum\ X-ray bursts from \src\ as a function of burst number, which
increases monotonically with time.
The horizontal solid
line shows the mean peak flux of the radius expansion bursts (diamonds),
while the
dashed lines show the $1\sigma$ limits.  Error bars indicate the $1\sigma$
uncertainties on each measurement.
  \label{fluxdist} }
\bigskip
\noindent
%%%%%%%%%%%%%%%%%%%%%%%%%%%%%%%
\centerline{\epsfxsize=8.5cm\epsfbox{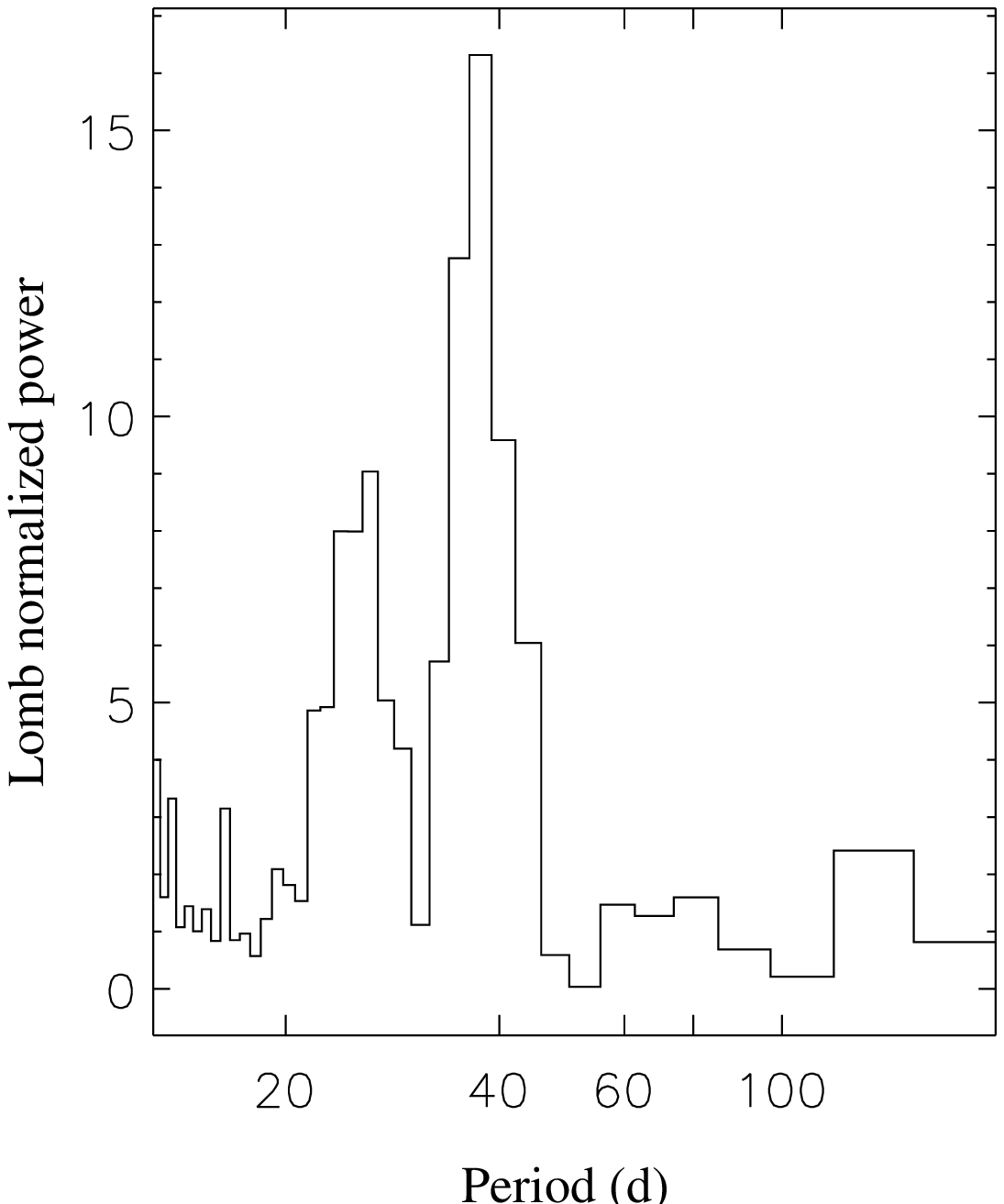}}
 \figcaption{
The Lomb-normalized periodogram of the time variation of the peak fluxes
\fpkre\/ from 50 radius-expansion bursts occurring before MJD~51500 (of
\preburst\/ radius expansion bursts in total).  Note the indications for
excess power between 30--60~d; the most significant peak is at 38~d, with
a Lomb power of 16.3.
 \label{peakflux} }
\bigskip
\noindent
%%%%%%%%%%%%%%%%%%%%%%%%%%%%%%%%%%%%
flux measurements.  Throughout this paper we quote the 
unadjusted \xte\/ peak
fluxes, but for distance estimates (see \S\ref{dist}) we consider
the possible effects of this systematic offset.

To measure the (pre-burst) persistent flux we estimated the instrumental
background using {\sc pcabackest} version 3.0, and fit the
background-subtracted pre-burst spectrum with an absorbed blackbody and
power law model.  The typical reduced-$\chi^2$ ($\chi^2_\nu$) was 1.2.
Gain-corrected X-ray colors were also calculated from the pre-burst
spectrum.  We determined correction factors for the count rates in the
various energy bands by comparison of Crab spectra over the lifetime of
the instrument \cite[see][for more details of the correction
method]{muno02}.  The soft color was calculated as the ratio between the
background-subtracted gain-corrected count rates in the 3.6--5.0~keV and
2.2--3.6~keV bands, and hard color as the ratio between counts in
8.6--18~keV and 5.0--8.6~keV bands.

\subsection{Peak burst fluxes}

We detected \burstnum\ X-ray bursts observed by \xte\/ from \src\ between
1996 February 15 and \lastpublic, with peak burst fluxes in the range
(0.3--$1.2)\times10^{-7}\ \epcs$.  We also detected four much fainter
bursts, which peaked at $\approx5\times10^{-9}\ \epcs$; two on 1996 May 3,
and one each on 2001 May 27 and May 29.  These burst times closely
followed active periods of the nearby source MXB~1730$-$33 (the ``Rapid
Burster''), just $0.53\arcdeg$ away, compared to field of view of \xte\/ of
$1\arcdeg$.  Since the next brightest burst from \src\ was a
factor of 6 brighter, we conclude that the four very faint bursts actually
originated with the Rapid Burster, and exclude them from our analysis.

Most of the X-ray bursts from \src\ exhibited some degree of radius
expansion.  As a working definition, we considered that
the local Eddington limit had been reached when (i) the blackbody
normalization $K_{\rm bb}$, which is directly related to the surface area
of the emitting region, reached a local maximum close to the time of peak
flux; (ii) lower values of the normalization $K_{\rm bb}$ were measured
following the maximum, with the decrease significant to $4\sigma$ or more;
and (iii) there was evidence of a significant (local) decrease in the
fitted temperature $T_{\rm bb}$ at the same time as the increase in the
normalization $K_{\rm bb}$.
For \preburstmaybe\ of the bursts there was weak evidence of
radius expansion, at only the 2--$3\sigma$ level.  However, the
distribution of peak fluxes in this class of bursts was consistent with
that of the bursts exhibiting more significant radius expansion, and so we
treated them as one population.

The peak bolometric fluxes \fpk$=\max(F_{{\rm bol},i})$ 
were bimodally distributed, with the radius-expansion bursts around a mean
of \fmean, and the non-radius expansion bursts
around \fnopremean\ (Fig.  \ref{fluxhist}).
The peak fluxes of the radius expansion bursts were roughly proportional
to the fluence, up to an asymptotic level of
$\approx1.2\times10^{-7}\ \epcs$ (Spearman's rank correlation $\rho=0.78$,
significant at approximately the $6\sigma$ confidence level). 
The burst fluence generally ranged from (0.04--$0.9)\times10^{-6}\ {\rm
erg\,cm^{-2}}$.
The non-radius expansion bursts had fluences below $0.5\times10^{-6}\ {\rm
erg\,cm^{-2}}$, while the majority of the radius expansion bursts had
fluences between (0.4--$0.8)\times10^{-6}\ {\rm erg\,cm^{-2}}$.
The brightest radius-expansion burst (\#72, on 2001 February 9 03:01.40
UT) also had the largest fluence. While its peak flux at
$(1.21\pm0.05)\times10^{-7}\ \epcs$ was around 30\% larger than the mean for
the remaining radius expansion bursts, the fluence was almost 150\%
greater than the mean at $(1.409\pm0.016)\times10^{-6}\ {\rm erg\,cm^{-2}}$. 

In the radius expansion bursts, the peak flux was generally reached
during the contraction stage following the radius maximum.
This is contrary to basic theory of Eddington limited bursts, which
predicts that the observed bolometric flux should be constant throughout
the radius expansion and contraction (once the flux is corrected for the
effect of gravitational redshift, which is naturally variable due to the
changing elevation of the photosphere above the neutron star surface).
Furthermore, as noted by \cite{vs01}, some bursts exhibited unusual
variation of the fitted radius as the burst evolved. In about half the
bursts the radius expansion and subsequent contraction were highly
significant, but following the radius minimum (which is usually assumed to
be the time when the expanding material ``touches down'' on the neutron
star surface) the apparent radius increased again, up to a level which in
some cases was as high as the initial radius maximum.
Since the initial rise and fall in the fitted radius generally exceeded our
criteria for classification as a radius expansion burst, we included these
unusual cases with the other ``normal'' radius expansion bursts. However,
there are clearly additional factors influencing
the observed spectra throughout the burst \cite[possibly involving changes
in the photospheric composition, perhaps related to ejection of an
hydrogen-rich envelope;][]{seh84}. While these additional effects may also
help to determine the peak flux of radius expansion bursts, in general
their greatest influence appears to be on the evolution later in the burst,
after the initial 5~s or so when the radius expansion takes place.

 \subsection{Peak flux variation in radius-expansion X-ray bursts}

The peak fluxes of the radius-expansion bursts from \src\ show a
significant ($\chi^2=1760$ for 60 degrees of freedom) deviation from a
constant value (Fig. \ref{fluxdist}).  The standard deviation of the
peak fluxes was \fstddev, corresponding to a fractional rms of
\fstddevp\%; the net variation was \netvarp\%.

While the bursts were clustered extremely irregularly in time (primarily
due to the irregular scheduling of pointed observations) the peak fluxes
of radius expansion bursts that were close in time gave a strong
suggestion of systematic evolution, on time scales of a few tens of days.
To illustrate this evolution, we calculated a Lomb-normalised periodogram
of the peak flux of the radius-expansion bursts.  Like the ASM
periodicity, the time scale of the variation in the peak fluxes did not
appear to be consistent over the full set of bursts we measured from
\xte\/ observations. Thus, we selected only those bursts before MJD 51500
to calculate the periodogram (note that this is also when the $\sim40$-d
periodicity in the ASM lightcurve became much weaker; Fig.
\ref{asmdat}). The periodogram shows evidence for
excess power between 30--60~d; the most significant peak was at 38~d, with
a Lomb power of 16.3 (estimated significance $>5\sigma$ from Monte-Carlo
simulations; Fig. \ref{peakflux}). We fit the peak fluxes folded on the
38~d period with a sine curve, with a resulting fractional rms amplitude
of 7.7\%.  The next most significant peak in the periodogram
was at 26~d, with power 9.0 ($3.1\sigma$).
Additional evidence that the changes in the peak burst
%%%%%%%%%%%%%%%%%%%%%%%%%%%%%%%%%%%%
\centerline{\epsfxsize=8.5cm\epsfbox{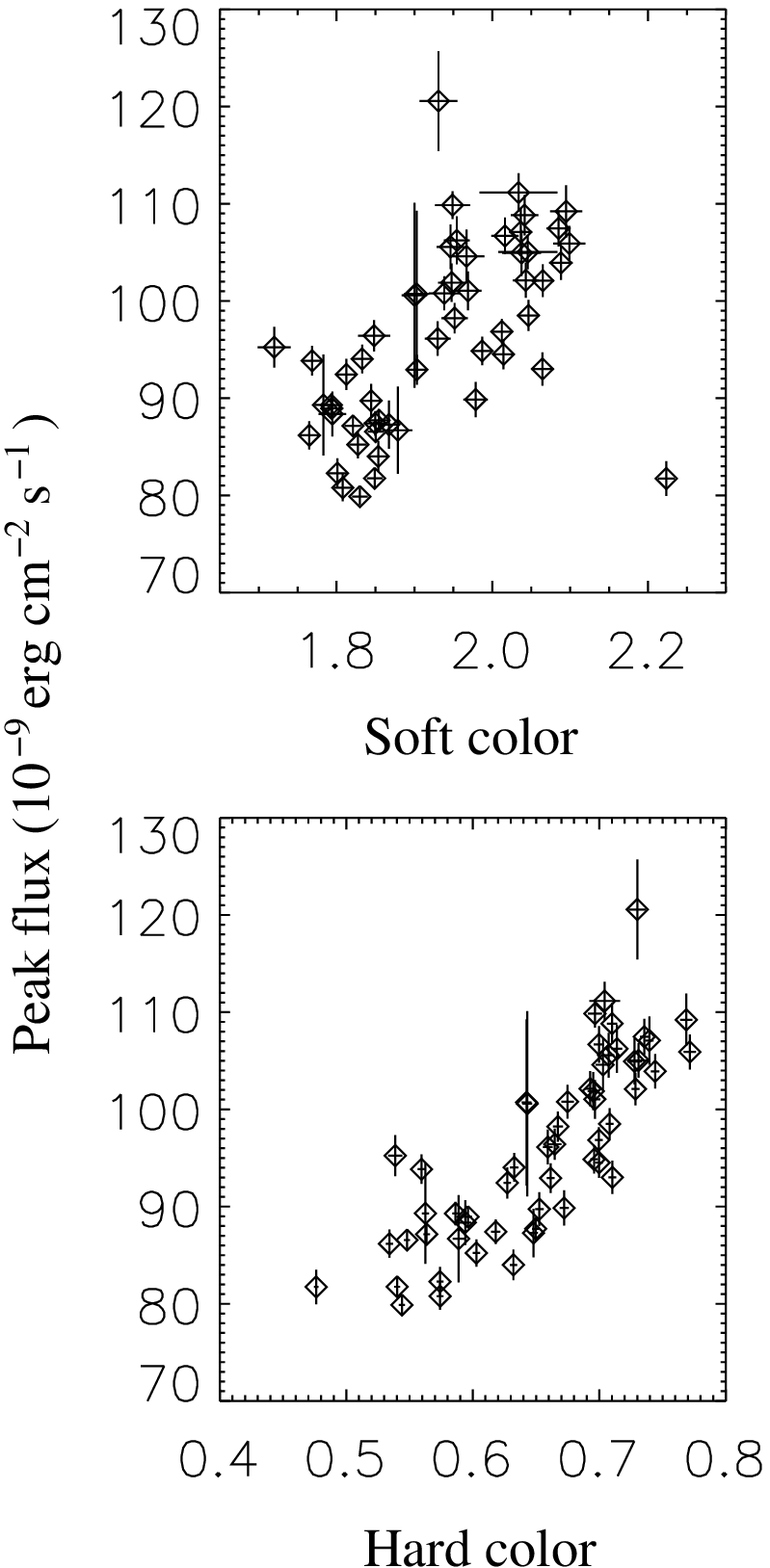}}
 \figcaption{Peak flux \fpkre\/ of radius-expansion thermonuclear
X-ray bursts from \src, as a function of the gain-corrected soft
(3.5--5.0~keV/2.2--3.6~keV) and hard (8.6--18~keV/5.0--8.6~keV) X-ray
colors of the persistent flux, prior to each burst. Error bars represent
the estimated $1\sigma$ uncertainties.
 \label{colors} }
\bigskip
\noindent
%%%%%%%%%%%%%%%%%%%%%%%%%%%%%%%%%%%%
flux was linked to
the long-term source evolution comes from significant correlations
measured between \fpkre\ and both the hard and soft gain-corrected colors
(Fig. \ref{colors}). The (Spearman's) rank correlation between the soft
color and \fpkre\ was $\rho=0.62$, with a significance of
$6.8\times10^{-6}$ (equivalent to $4.5\sigma$); for the hard color,
$\rho=0.84$, with a significance of $3.4\times10^{-15}$ (equivalent to
$6.1\sigma$).

While the 38~d periodicity found in the radius-expansion burst peak fluxes
was corroborated by a significant detection at a similar period in the ASM
flux history, the 26~d period was not. Thus, we consider the evidence for
the presence of the latter signal to be weak.
The apparent quasi-periodic behavior of the ASM flux modulation,
coupled with the similarity in the time scales and
%%%%%%%%%%%%%%%%%%%%%%%%%%%%%%%%%%
\centerline{\epsfxsize=8.5cm\epsfbox{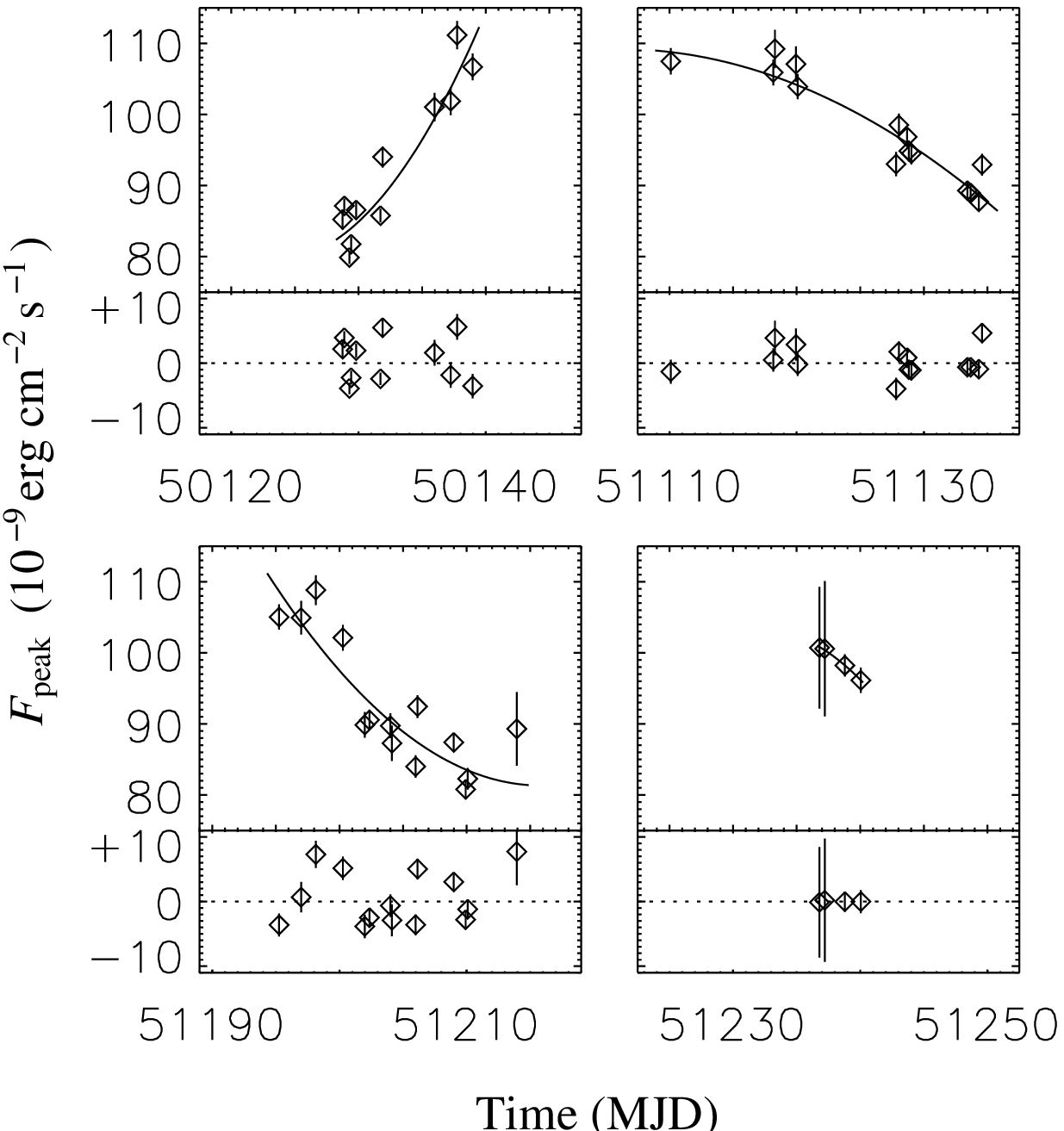}}
 \figcaption{ The peak burst fluxes \fpkre\ of selected
 subsets of radius-expansion bursts from \src, with the best-fit
 second-order polynomial overplotted as a solid line.  The lower part of
 each panel shows the residuals after the polynomial fit is subtracted. Error
 bars show the $1\sigma$ uncertainties.
 \label{subint} }
\bigskip

\centerline{\epsfxsize=8.5cm\epsfbox{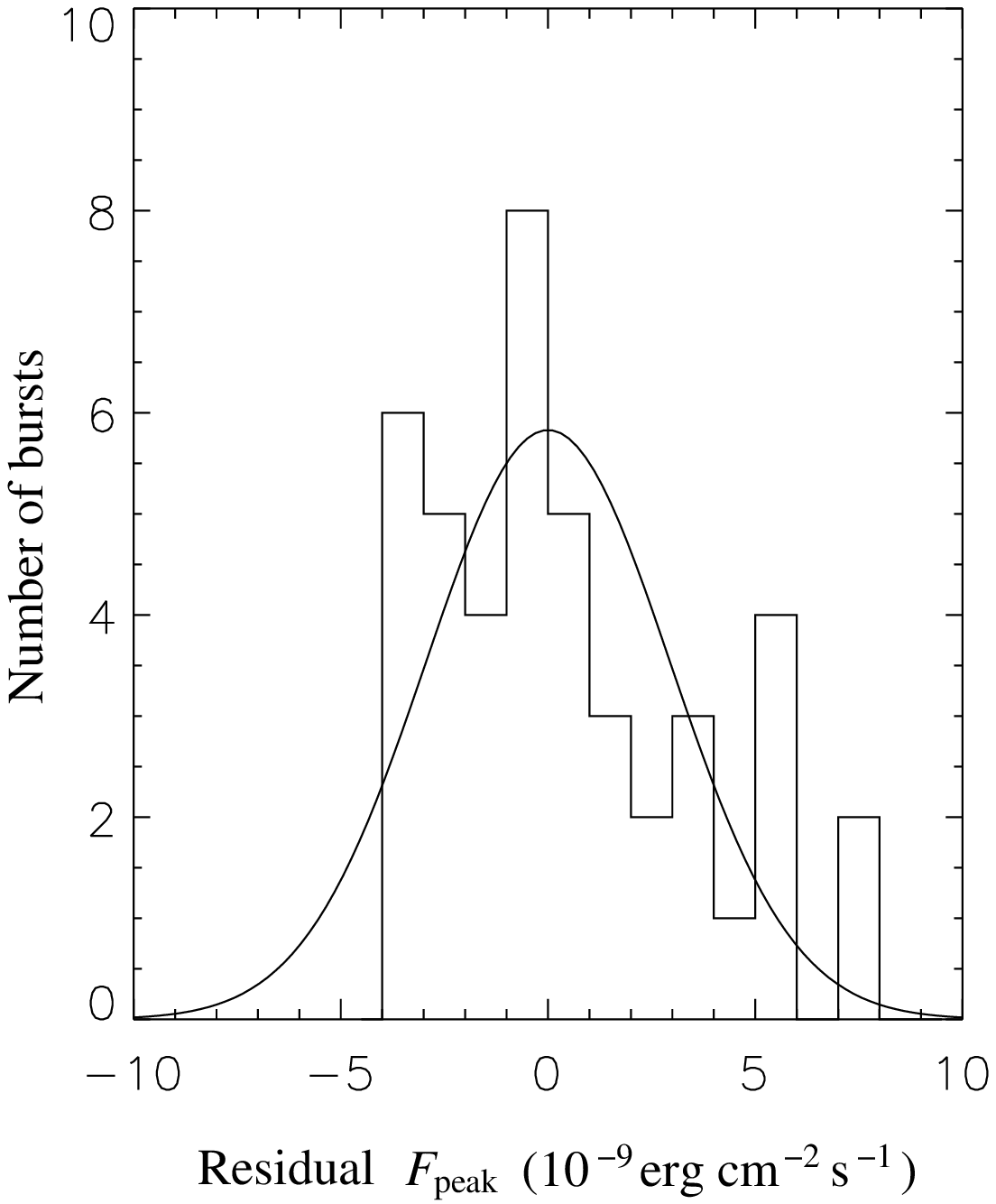}}
 \figcaption{The combined
 distribution of residuals from the fits shown in Fig. \ref{subint},
 with the expected Gaussian distribution for the measured
 $\sigma=2.8$\% overplotted.
 \label{fluxresid} }
\bigskip
\noindent
%%%%%%%%%%%%%%%%%%%%%%%%%%%%%%%%%%
modulation amplitudes of the ASM flux and the radius-expansion 
burst peak fluxes, suggest
that
these two phenomena share a common cause.
The uneven time sampling of \fpkre\ and the time-dependence of the ASM
periodicity (Fig. \ref{asmdat}) did not allow us to test this hypothesis
for the entire set of radius-expansion bursts.  However, we have
identified four intervals, in which multiple radius expansion bursts were
detected over time spans of no more than $\approx 30$~d.  The resulting 4
subsamples contain a total of 41 of the \preburst\ radius-expansion bursts
measured from \src.  For each of these intervals we attempted to remove
the long-term variation
by fitting the overall \fpkre\ time-dependence using a second order
polynomial (Fig.~\ref{subint}). The combined distribution of the
residuals from each of the polynomial fits is shown in
Fig.~\ref{fluxresid}. The distribution is consistent with Gaussian noise
with $\sigma=3.2$\% of the mean peak flux, compared to the averaged $1\sigma$
flux measurement errors of 2.1\%.

\section{DISCUSSION}
\label{disc}

We have studied \burstnum\ thermonuclear X-ray bursts from \src, which
includes \preburst\ exhibiting evidence for photospheric radius expansion,
as observed by the {\em RXTE}/PCA.
We have shown that the radius expansion bursts exhibit a significant
variation in their peak fluxes \fpkre.  The fractional standard deviation
was $\simeq \fstddevp$\%, while the total variation was
$\simeq$\netvarp\%.  This is significantly larger than the formal
measurement errors, which were typically $\simeq 2$\%.  This result appears
inconsistent with the simple picture of radius-expansion bursts
\cite[e.g.][]{lew93}, according to which the peak flux should be nearly
constant and equal to the Eddington critical flux.  However, we must also
consider the possibility of a variety of systematic effects arising as a
consequence of our analysis method, which may contribute (or give rise) to
the measured variation.

\subsection{Possible systematic effects}
\label{system}

While subtraction of the pre-burst emission as background has dubious
theoretical justification, and may have originally been adopted for convenience,
the method has been shown in several analyses to be relatively robust
\cite[e.g.][]{vpl86,kuul01}. An implicit assumption is that the persistent
emission remains unchanged throught the burst. This may not be true,
especially for the very energetic, radius-expansion bursts that may
disrupt the inner accretion flow.  As a result, variations in the true
persistent flux $F_{\rm per}(t)$ during the burst may contribute to
variation in the measured bolometric burst flux.  The persistent flux
prior to the radius expansion bursts in \src\ was typically around
$3.7\times10^{-9}\ \epcs$, but for some bursts was as high as
$5\times10^{-9}\ \epcs$.  The net variation over all the bursts was only
$3.2\times10^{-9}\ \epcs$, which is insufficient to account for the
\netvar\ net variation in the peak burst fluxes.  Even if the persistent
flux was quenched completely during the burst, this would still only give a
variation of at most $5\times10^{-9}\ \epcs$.
Thus, variations in the peristent flux cannot completely account for the
observed variation in \fpkre, but may contribute to some extent.
If this is the case, we may expect a correlation between the
\fpkre\ and the pre-burst persistent flux.  In fact, \fpkre\ and the
persistent flux were 
weakly anticorrelated, with Spearman's rank
correlation
$\rho=-0.32$ (estimated signifcance $1.5\times10^{-2}$, equivalent to
$2.4\sigma$). If anything, the persistent flux variation serves to
suppress slightly the true variation in \fpkre.

The
bolometric correction (equation \ref{flux}) typically adds $\simeq 7$\% to
the peak 2.5--20~keV flux of the radius expansion bursts as measured by
the PCA.  This correction may influence our result in one of two ways.  On
the one hand, systematically biased bolometric corrections may give rise
to a variation in flux which is not present in the observed source fluxes
(restricted to the PCA passband).  We can easily rule out this
possibility, since the flux integrated just over the PCA passband exhibits
identical (fractional) variation as the inferred bolometric flux.
Alternatively, deviations in the spectrum from a perfect blackbody may
give rise to (unobserved) variations in the flux outside the PCA passband
that contribute to the true bolometric flux variation. In the latter case
these unobserved bolometric flux variations may (for example) partially or
completely compensate for the variation we see in the PCA passband, so
that the bolometric flux variation we infer is exaggerated, or erroneous.
The second effect is more difficult to discount, since (obviously) the
spectral variations outside the PCA passband are not measurable with the
present data.  However, the $\simeq 7$\% typically added to the flux in
the PCA passband by the bolometric correction is significantly smaller
than the $\simeq$\netvarp\% observed variation in the peak fluxes. In
order to compensate for the flux variation we observe, this contribution
would then need to fluctuate by a factor of $\simeq6$, which seems unlikely.

Thus, we conclude that the observed variation was not an
artifact of any aspect of our analysis method and, hence, must be genuine.  

\subsection{Origin of the peak flux variation}
\label{opfv}

The large number of radius-expansion bursts observed from \src\ with
{\em RXTE}/PCA
along with the long-term flux history accumulated by the ASM
provides for the first time a plausible cause for the observed broad
distribution of peak burst fluxes. The fact that (i) the peak burst
fluxes were correlated with the X-ray colors of the persistent emission,
(ii) they varied in a quasi-periodic manner, and (iii) the timescale and
fractional amplitude of the variability were similar to those of the
persistent emission, strongly suggest that the same phenomenon causes
the variability in both the persistent and burst fluxes.
 
Since the variability of the persistent emission is not coherent, it is
unlikely to be due to orbital modulation. As noted by \cite{kong98}, its
timescale (whether $\sim 30$ or 60--70~d) is much longer than would
normally be expected for the orbital period of a Roche lobe-filling
low-mass X-ray binary.  Such ``super-orbital'' periodicities are observed in
several similar sources \cite[e.g.][]{white95}, and are generally
attributed to variations in the accretion geometry, possibly caused by the
precession of a warped accretion disk about the neutron star.  If the
persistent emission is modulated at this timescale because of a slowly
evolving warp in the accretion disk that is reflecting a small fraction of
the X-ray luminosity of the central object to the observer, then the peak
burst fluxes would also be modulated at the same timescale and with
a similar amplitude. Our analysis of the radius-expansion 
bursts of \src\
strongly suggest that this is the case.
Further support is provided by the highly significant correlation
observed between the peak flux and the fluence for the
radius-expansion bursts. If there was no additional contribution to the
burst flux from disk reflection, theory predicts that the peak flux would
be independent of the fluence. 

It remains to be established whether, given the conditions in the
\src\ system, the accretion disk can become warped, undergo precession at
approximately the measured (quasi-) periodicities, and give rise to the
observed degree of modulation of the persistent and peak radius-expansion
burst fluxes.  The disk warp may arise (for example) from non-axisymmetric
radiation pressure forces \cite[e.g.][]{pringle96,mbp96,mb97}.  The
conditions required for the initial warping, and steady precession
thereafter, depend primarily upon the orbital separation and the
efficiency of accretion \cite[]{od01}. The orbital parameters are
presently unknown for \src, but are in principle measurable; the accretion
efficiency is much more difficult to measure for this, or any other, LMXB.
Thus, our present level of knowledge is not sufficient to reject the
hypothesis that a precessing, warped disk (whether arising by radiation
instabilities, or some other mechanism) is present in \src.
While the question of whether a disk warp can give rise to the observed
modulation is comparatively more straightforward, we are still limited by
the lack of measured system parameters, in particular the inclination.
For a flat accretion disk \cite{ls85} calculated an anisotropy factor of
2.8, depending upon the inclination angle. The precessing of a warped disk
is likely to affect the proportion of reprocessed radiation observed
during the burst in the same way that varying the inclination would. 
The derived anisotropy factor is more than sufficient to explain the
observed modulation in the peak flux of radius-expansion bursts.

Because of these uncertainties, we can most likely adopt a relatively wide
range of parameters (disk warping angle, disk albedo) which will give rise
to a modulation of at least the level measured in \src; thus, such an
approach would also have no ability to rule out warped disk precession as a
mechanism for the X-ray flux modulation.
We note that in the archetypical precessing warped disk system
Her~X-1, periodic obscuration of the neutron star by the disk gives rise
to a modulation of the persistent X-ray flux of essentially 100\%
\cite[e.g][]{slw00}, much larger than the $\sim10$\% measured for \src.  Since
it exhibits neither X-ray eclipses or dips, \src\ must have a lower
inclination ($i\la85\arcdeg$) than Her~X-1, making obscuration by the disk
less likely.  For a disk warped to the degree inferred for Her~X-1
($20\arcdeg$ at the outer edge), the {\it a priori} probability for
obscuration in \src\ is $\sim25$\%.
Even if the inclination is not sufficiently high to permit obscuration,
X-ray reflection from the disk, coupled with the variations in the
projected disk area due to the precessing warp, may yet be sufficient to
give rise to the observed modulation.

\subsection{Possible anisotropy of the burst emission}
\label{aniso}

When the systematic trends in the variation of the peak burst fluxes are
removed, the residual variation is only 2.8--3.2\%, which is
comparable to the typical measurement uncertainty of 2\%. This has a very
important implication for the anisotropy of radius-expansion bursts
in \src, as described by the parameter $\xi$ in equation~(\ref{ledd}). 
The small residual scatter of the peak fluxes strongly suggests that the
intrinsic variation of the peak burst flux is also small, $\simeq1$--2\%.
It seems unlikely that we observe the same face of the neutron star at the
same orientation during every one of these bursts, particularly given the
rapid rotation inferred from the burst oscillations
\cite[364~Hz;][]{stroh96}. Additionally, these same oscillations are
almost never observed during the radius expansion episode itself, even if
they are present earlier or later in the burst \cite[]{muno02b}. We
conclude that the longitudinal dependence of the burst flux during the
radius expansion episodes is negligible. A latitudinal variation in flux
remains plausible, particularly since the effective gravity is smaller at
the neutron star equator than at the poles, and so we might expect a
greater degree of expansion of the atmosphere there.
However, we observe significant variation in the blackbody normalization
when the peak burst flux is achieved, which suggests that the radius
expansion episodes reach different peak radii. Since we might expect the
degree of latitudinal anisotropy to vary with increasing radius, the
effect of such a latitudinal variation of flux would be a dependence of
the peak flux on the blackbody normalisation at the peak, which is not
observed. Thus, we conclude that the degree of latitudinal flux anisotropy
is most likely also limited by the (inferred) intrinsic variation of the
peak burst fluxes.  We conclude that the burst emission during the radius
expansion episode is isotropic to within $\simeq1$--2\%. 
Note that it is still possible for the burst emission at the neutron star
surface to be significantly anisotropic, but that this anisotropy is smoothed
out through reprocessing in the extended atmosphere present during the
radius expansion episodes.

\subsection{Consequences for distance estimates}
\label{dist}

Studies such as this provide a measure of the systematic uncertainties of
the distance estimates of X-ray sources that are based solely on
Eddington-limited bursts \cite[see, e.g.,][]{vpw95}.
We note that the standard deviation we measure is within the typical peak
burst flux uncertainty ($\simeq15$\%) measured by \cite{kuul02} for the
globular cluster burst sources.  While the inferred (intrinsic) isotropy
of the burst radiation (section \S\ref{aniso}) allows us to at least
eliminate that contribution to uncertainties in distance estimates, the
additional systematic error contributed by the observed scatter in the
peak burst fluxes is still smaller than the usual other uncertainties due
to the unknown neutron star mass and atmospheric composition.
Furthermore, without a detailed understanding of the degree of
reprocessing occurring in the region around the neutron star, we cannot at
this time determine the intrinsic peak luminosity of the radius expansion
bursts, from the broad distribution we have observed.

Nevertheless, we now
calculate a probable range for the distance to \src, given plausible
values for the neutron star mass and atmospheric composition.
We identify the minimum peak flux of the radius expansion bursts as the
best estimate of the Eddington limit;
this burst will have the smallest contribution due to reprocessed radiation,
and thus will provide the best estimate of the intrinsic maximum flux.
Since the peak flux is typically reached near
the end of the radius contraction, we calculate the gravitational redshift
parameter at the neutron star radius $R_{\rm NS}=10$~km. We also reduce
our observed fluxes by 20\% to correct for the observed systematic flux
offset measured for \xte\/ (see \S\ref{pca}), so that the inferred Eddington flux
is $6.2\times10^{-8}\ \epcs$.
Thus, for a 1.4(2.0)~$M_\sun$ neutron star with cosmic atmospheric abundance 
($X=0.7$), the distance is
4.4(4.8)~kpc.
For a pure He atmosphere the distance is 30\% greater.
These values are roughly consistent with previous estimates
\cite[]{vp78,bas84,kam89} and place the source within 12~pc of the
Galactic plane, about 4~kpc from the center.

\acknowledgments

This research has made use of data obtained through the High Energy
Astrophysics Science Archive Research Center Online Service, provided by
the NASA/Goddard Space Flight Center.  We thank Mike Nowak, Fred Lamb and
Erik Kuulkers for their helpful comments on the paper. This work was
supported in part by the NASA Long Term Space Astrophysics program under
grant NAG 5-9184.

\end{document}